\documentclass[]{jfm}

\usepackage{newtxtext}
\usepackage{newtxmath}
\usepackage{natbib}
\usepackage{hyperref}
\usepackage{float}

\hypersetup{
    colorlinks = true,
    urlcolor   = blue,
    citecolor  = black,
}

\newcommand{\RomanNumeralCaps}[1]

\title{Stages of turbulence generation and decay in a T-shaped mixer}

\author{Mohammad Mehdi Zamani Asl\aff{1}
  \corresp{\email{mehdi.zamani@zarm.uni-bremen.de}},
  Marc Avila\aff{1,2}
 }

\affiliation{\aff{1}Center of Applied Space Technology and Microgravity (ZARM), University of Bremen, Am Fallturm 2, Bremen 28359, Germany
\aff{2}MAPEXCenter for Materials and Processes, University of Bremen, Am Biologischen Garten 2, Bremen 28359, Germany}

\begin{document}
\maketitle

\begin{abstract}
The T-shaped mixer is widely used in fundamental studies of chemical engineering. Its transitional regime is well understood, whereas the turbulent dynamics has received scarce attention so far. Here we perform direct numerical simulations of the turbulent regime for Reynolds numbers up to $Re=2000$ at Schmidt number $Sc=1$. Our analysis reveals two distinct stages along the mixing channel prior to relaxation toward duct flow. Near the junction, a jet-like flow forms and exhibits the approximately self-similar behaviour of transitional planar jets. Subsequently, a decay region characterised by power-law decay of turbulent kinetic energy, dissipation and scalar variance emerges. For the velocity field, the observed exponents are consistent with those of decaying turbulence in bounded domains, whereas the scalar-variance exponent is consistent with that of unbounded turbulence. We argue that this apparent discrepancy is a consequence of the mixing process progressing from the center of the channel toward the side walls in the decay region, while turbulence already fills the channel cross-section entirely at the end of the jet region.The time-averaged mixing state presents error-function profiles of the scalar in the transverse direction, similar to the laminar cases, and is quantified here through a stream-wise evolving effective diffusion coefficient.
\end{abstract}


\section{Introduction}
\label{sec:headings}

The mixing of fluids is a common process in nature and industry. The instantaneous mixing rate is controlled by the contact area between the fluids and the molecular diffusion coefficient, $D$. In practice,  mixing devices are devised to generate highly turbulent flows, where multi-scale eddies greatly increase the contact area \citep{Dimotakis}. The T-shaped mixer is widely used because of its simple design and mixing efficiency \citep[see e.g.][for recent reviews]{Camarri2020,LI2023}. Despite its geometric simplicity, it displays a remarkably rich sequence of flow regimes as $Re$ raises, ranging from symmetric laminar flow to the development of engulfment patterns and time-periodic symmetric oscillations \citep{hoffmann2006experimental,Ameel2010,Mariotti2018}. Chaotic motion arises at $Re\sim500$ and the flow gradually turns turbulent as $Re$ is further increased \citep{Tobias2017}. The Reynolds number is defined as $Re=\frac{Ud}{\nu}$, where $U$ is the mean inlet velocity, $d$ the hydraulic diameter of the inlet and $\nu$ the kinematic viscosity of the fluid. In the specific geometry considered here, the outlet channel has aspect ratio $2$ and the mean outlet velocity is also $U$. 

Despite its prevalence in chemical engineering \citep{schwarzer2006predictive}, the turbulent regime of T-mixers has been largely unexplored until recently. \citet{Tobias2019} studied the influence of the inlet conditions on the turbulent mixing for $Re \leq 4000$. They showed that by carefully manipulating the state of the inlet, the mixing rate could be greatly enhanced. For laminar inlets, they found that the turbulent statistics became $Re$-independent at $Re\gtrsim 2000$ already. Recently, \citet{Huixin2025} experimentally investigated mixing in a T-mixer with quasi-laminar inlets using PIV and PLIF up to $Re=5000$. They measured turbulent kinetic energy and scalar variance at three cross-sections along the outlet and provided evidence for the scaling of the scalar variance in the viscous-convective range predicted by \citet{Batchelor1959} in the high-$Sc$ limit. 

Despite this recent progress, the mechanisms of turbulence generation and decay in T-mixers, and their impact on mixing, remain poorly understood. We here fill this gap and examine the coherent structures and scaling of the turbulent quantities along the outlet up to $Re = 2000$ and laminar inlets. We relate our findings to established results of canonical turbulent flows, thereby shedding light on the physical mechanisms of turbulence in T-mixers. 


\section{Methods}
\label{sec:headings}

\subsection{Governing equations}
The nondimensional incompressible Navier-Stokes and scalar transport equations were solved to compute the fluid motion and mixing,
\begin{equation}
\frac{\partial\textbf{u}}{\partial t} + \textbf{u}\cdot\nabla\textbf{u} = -\nabla p + \frac{1}{Re}\Delta\textbf{u},\quad {\nabla}\cdot\textbf{u}=0,  
\end{equation} 
\begin{equation}
\frac{\partial{\theta}}{\partial t} + {\textbf{u}}\cdot\nabla{\theta} = \frac{1}{ReSc}\Delta{\theta}.
\end{equation}

\begin{figure}
  \centerline{\includegraphics{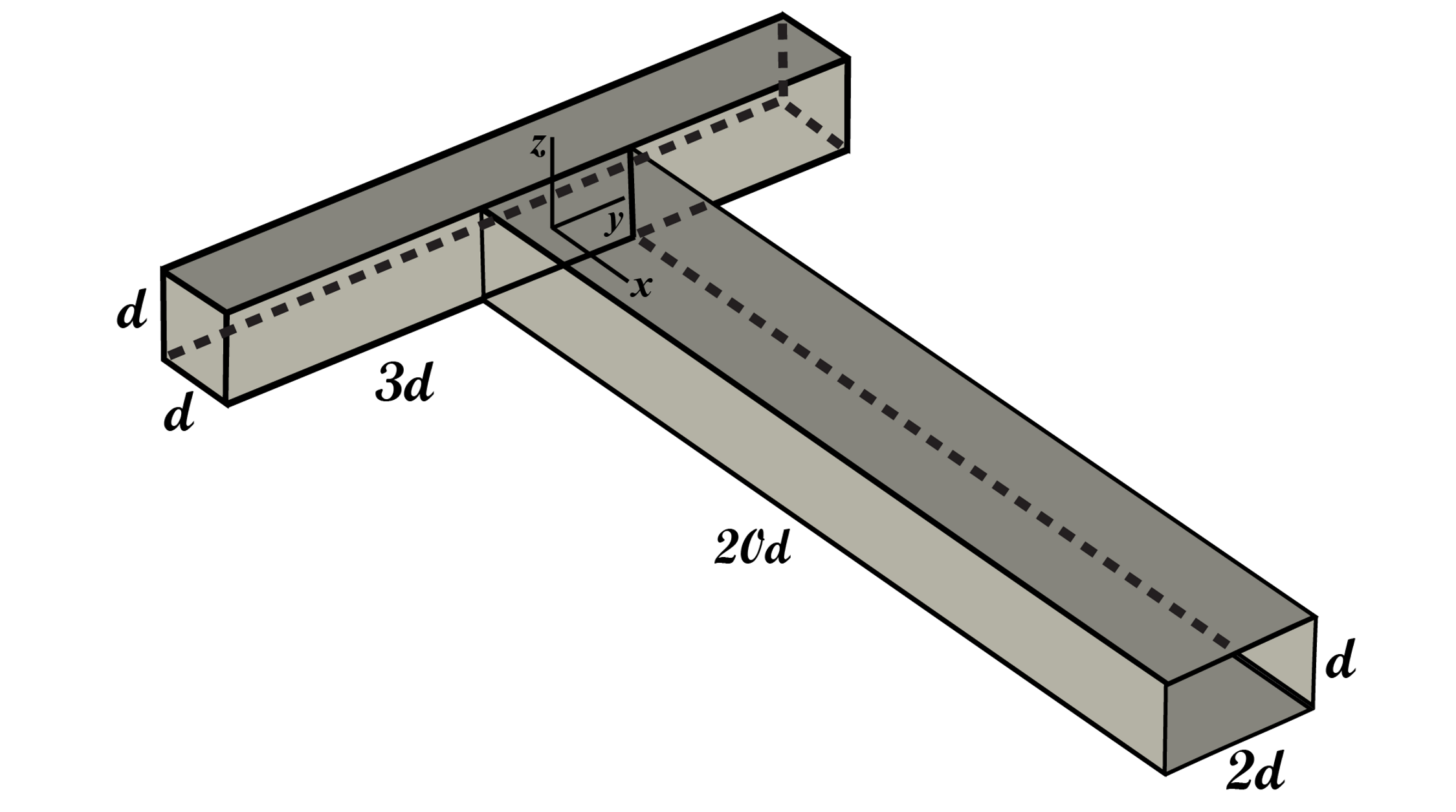}}
  \caption{Schematic of the computational domain.}
\label{fig:Schem}
\end{figure}
Figure \ref{fig:Schem} illustrates the geometry and dimensions of the simulated T-mixer. We use Cartesian coordinates, with the origin positioned at the center of the junction. The $y$-axis is aligned with the opposing square duct inlet, while the $x$-axis is oriented along the streamwise direction of the rectangular outlet duct. Lengths were nondimensionalized with the hydraulic-diameter of the inlets, $d$, and velocities with the bulk velocity, $U$. The Reynolds number in eqs. (2) and (3) is defined as  $Re=\frac{Ud}{\nu}$, where $U$ is the mean inlet velocity, $d$ is the hydraulic diameter of the inlet, and $\nu$ is the kinematic viscosity of the fluid. The Schmidt number $Sc=\frac{\nu}{D}$, where D is the molecular diffusion coefficient, was set to $Sc=1$.

\subsection{Boundary conditions}
\begin{figure}
  \centerline{\includegraphics{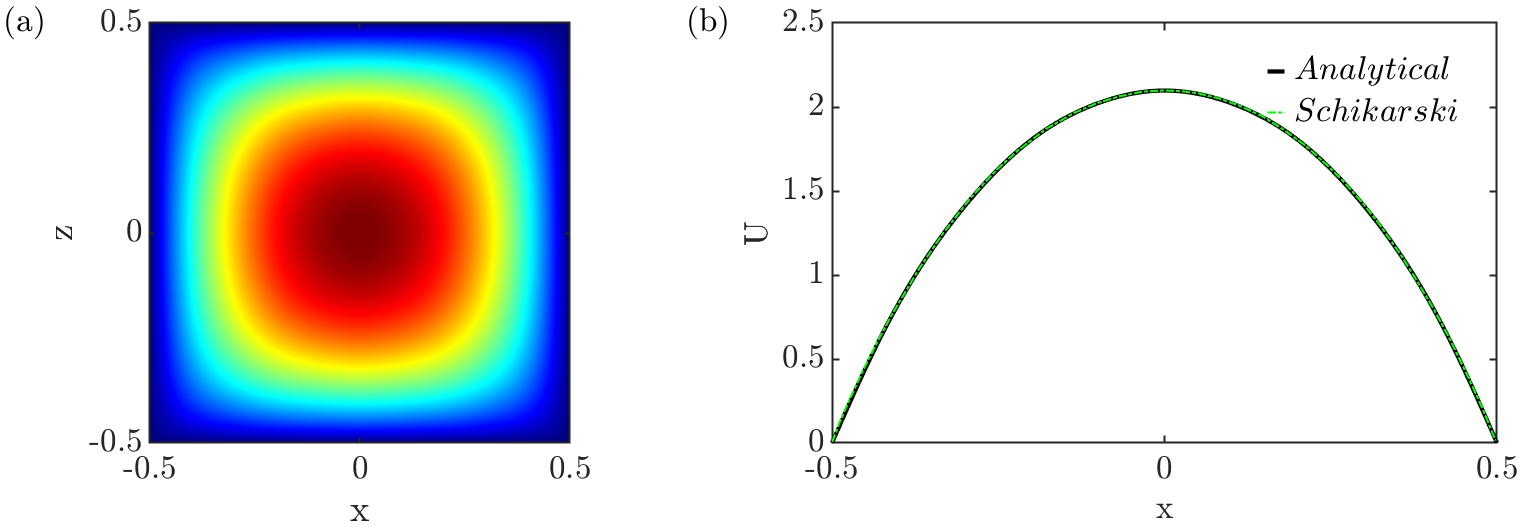}}
  \caption{Fully developed flow in a square duct was used as boundary condition at the inlets of the T-mixer (a) colormap of the stream-wise velocity (b) velocity profile at mid height ($z=0$). The dashed green line is from \citet{Tobias2019}.}
\label{fig:inlet_profile}
\end{figure}

At the inlets, a fully developed laminar velocity profile for a square duct \citep{London} was imposed as the boundary condition. In dimensionless form, it reads,
\begin{equation}\label{sec:shah}
u(x,z)=-\frac{4c_{1}}{\pi^3}\sum_{n=1,3,...}^{\infty}\frac{1}{n^3}(-1)^{\frac{n-1}{2}}\left[ 1-\frac{\cosh(n\pi x)}{\cosh(\frac{n\pi}{2})} \right]\cos(n\pi z),
\end{equation}
where the constant
\begin{equation}\label{sec:shahmean}
    c_{1}=-\frac{12}{\left[ 1-\frac{192}{\pi^5}\sum_{n=1,3,...}^{\infty }\frac{1}{n^5}\tanh(\frac{n\pi}{2}) \right]}
\end{equation}
ensures that $\int_{-\frac{1}{2}}^{\frac{1}{2}}\int_{-\frac{1}{2}}^{\frac{1}{2}} u(x,z)dxdz=1$.
A colormap of the velocity (a) and profile at $z=0$ (b) obtained by truncation of $n=5$ are shown in figure \ref{fig:inlet_profile}.
A stabilised outflow boundary condition was applied at the outlet to prevent any backflow into the domain \citep{DONG201483}. The no-slip boundary condition was applied at the walls. The scalar concentration was set to $\theta=1$ at the left inlet and $\theta=0$ at the right inlet, whereas homogeneous Neumann conditions, $\frac{\partial \theta }{\partial n}=0$, were applied at the walls and outlet. To compare the T-mixer outlet with a fully developed duct flow, we additionally performed simulations of a rectangular duct at $Re=2000$ with aspect ratio $2$ using periodic boundary conditions in the stream-wise direction (of length $20$).

\subsection{DNS}
 Nek5000, a massively parallel spectral-element code \citep{fischer2008nek5000}, was used to discretize the governing equations and compute the velocity field $(u,v,w)$ and the scalar $\theta$ through direct numerical simulation. Table \ref{tab:Sim} contains the detailed information of the simulated cases. For each case, the number of elements was selected according to the Kolmogorov scale estimated from the dissipation at the same Reynolds number in similar studies \citep{Tobias2017,Tobias2019}. All simulations were performed using the $\mathbb{P}_{N}-\mathbb{P}_{N}$ formulation in Nek5000, employing a polynomial order of $N-1=7$ for both velocity and pressure fields. Second-order implicit backward differentiation (BDF2) was used for time integration. A variable time-step was used to maintain the CFL number below $0.5$.

\begin{figure}
  \centerline{\includegraphics{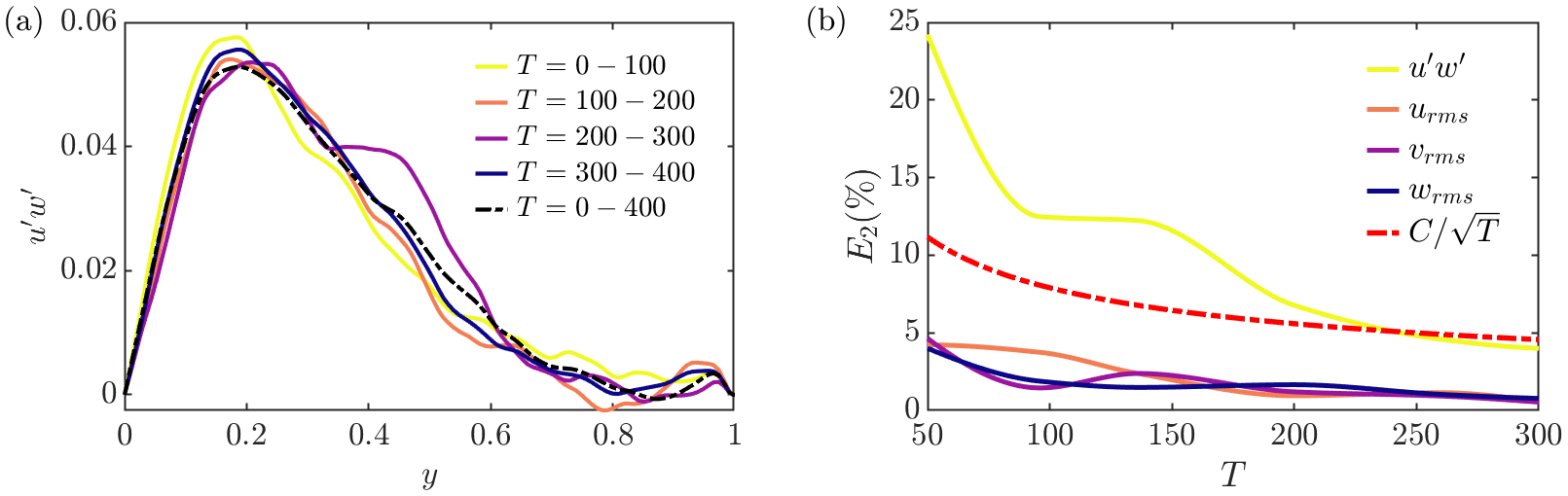}}
  \caption{(a) Reynolds shear-stress profiles averaged over successive 100 advective time unit intervals and over the full simulation time of 400 advective time units. (b) Relative errors of the Reynolds shear stress and r.m.s.\ of the velocity fluctuation as function of the averaging time, $T$, using the 400-unit average as reference. The red dashed line is a $C/\sqrt{T}$ fit to the data for $T\ge200$.}
\label{fig:convergence}
\end{figure}

\begin{table}
  \begin{center}
\def~{\hphantom{0}}
  \begin{tabular}{llllll}
        & $Re=700$  & $Re=1000$ &   $Re=1500$ & $Re=2000$ & $Duct$ \\[3pt]
       Number of elements   & 114300 & 152592 & 184680 & 370706 & 317400 \\
  \end{tabular}
  \caption{Number of elements used in the direct numerical simulations. The inlet conditions are laminar and periodic for the duct. The polynomial order is $N-1=7$.}
  \label{tab:Sim}
  \end{center}
\end{table}
The simulations were conducted in the Fritz CPU cluster of the Erlangen National High Performance Computing Center, utilising up to $4608$ computing cores. They were run for a total of 450 advective units at each $Re$. An initial phase of $50$ units was excluded from the data to ensure that the flow reached a statistically steady state. Following that, data were collected over an initial phase of $100$ units to converge the mean velocity field. Subsequently, other statistical quantities were computed by averaging over remaining $300$ units. 
\\
To assess the statistical convergence of the higher order quantities, we extended the averaging interval to 400 advective time units and computed the r.m.s.\ velocity fluctuations and Reynolds shear stress for $Re=2000$. In figure~\ref{fig:convergence}(a), we show the Reynolds shear-stress profiles averaged over successive 100-unit intervals and the full average. These results demonstrates that averaging over 100 time units is sufficient to accurately capture central region of the channel. In figure~\ref{fig:convergence}(b), we plot the relative error of the Reynolds shear-stress and velocity-fluctuation profiles as a function of averaging time, $T\leq 300$, and using the full average as a reference,
\begin{equation}
E_2^{(q)}(T)
=
100\,
\frac{
\left\|q_T(y)
-q_{400}(y)\right\|_2
}{
\left\|q_{400}(y)\right\|_2
},
\label{eq:statistical_convergence}
\end{equation}
where $q_{T}$ denotes the profile averaged over $T$ advective time units and the norm is evaluated over the transverse coordinate, $y$. All errors decrease with averaging time, broadly following the expected trend, $E_2^{(q)}(T)\propto T^{-1/2}$, at sufficiently long times. At $t=300$, the error is 4\% for $u'w'$ and below 1\% for all r.m.s.\ velocity fluctuations.

\section{Results}

\begin{figure}
  \centerline{\includegraphics{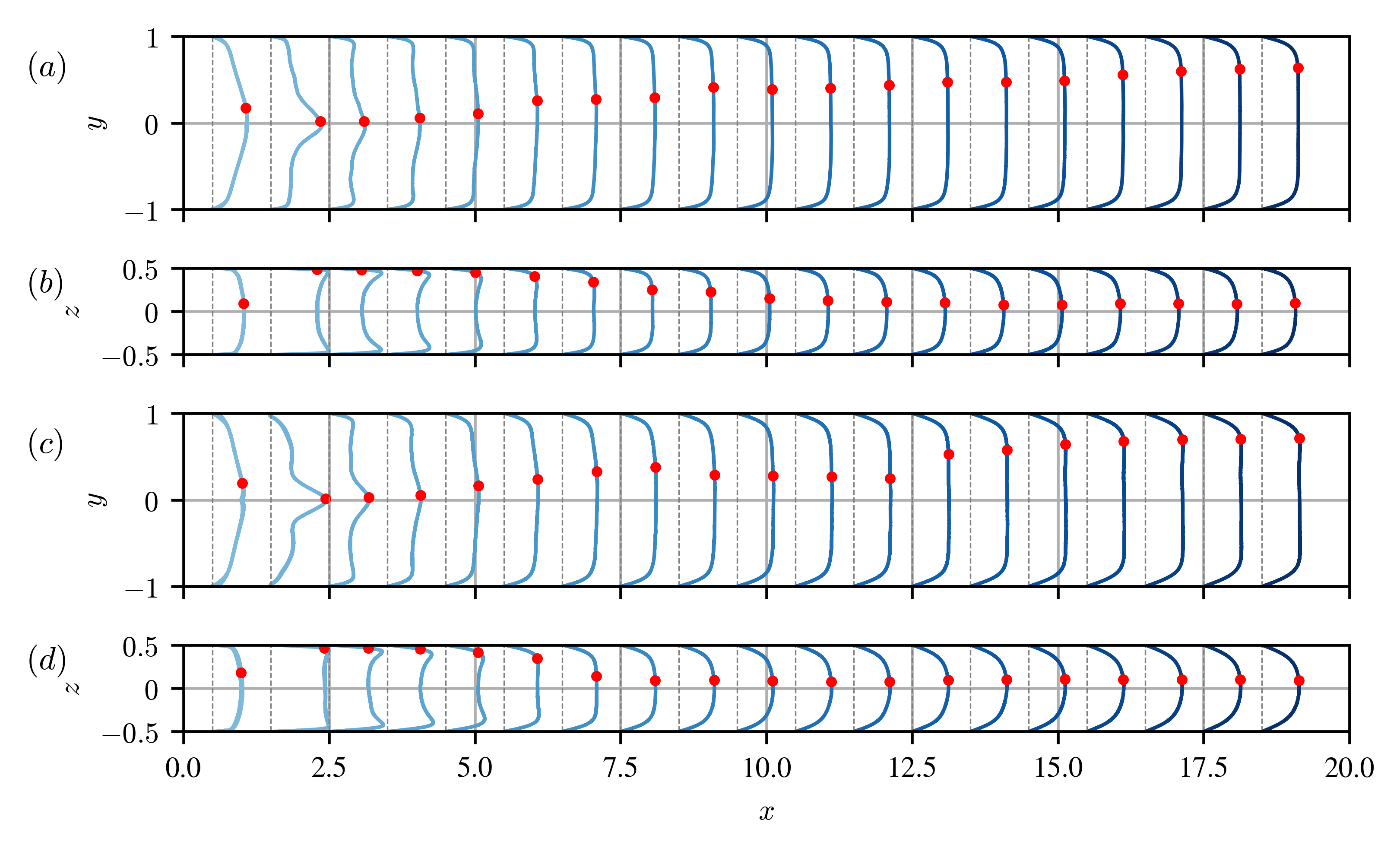}}
  \caption{(a)-(c) Time-averaged stream-wise velocity profiles at horizontal and (b)-(d) vertical mid-lines along the outlet at $Re=2000$ and $Re=700$, respectively. The red dots indicate 99\% of the centreline velocity and illustrate the boundary-layer development.}
\label{fig:vel_profile}
\end{figure}

In figure~\ref{fig:vel_profile}, we show time-averaged stream-wise velocity profiles at horizontal and vertical lines and various stream-wise stations along the outlet channel for (a)-(b) $Re=2000$ and (c)-(d) $Re=700$ cases.  After the junction ($x>0$), a prominent jet-like velocity profile is observed in the horizontal (span-wise) direction, characterised by a high-velocity core and steep near-wall gradients (figure~\ref{fig:vel_profile}(a) and (c)). In the vertical (wall-normal) direction, the profile is approximately constant except for strong near-wall shear layers (figure~\ref{fig:vel_profile}(b) and (d)). Hence, near the junction the velocity profile resembles that of a planar jet. After the jet widens toward the wall, the velocity profile becomes flat in both directions, which suggests a homogeneous, decaying turbulent state.

\begin{figure}
  \centerline{\includegraphics{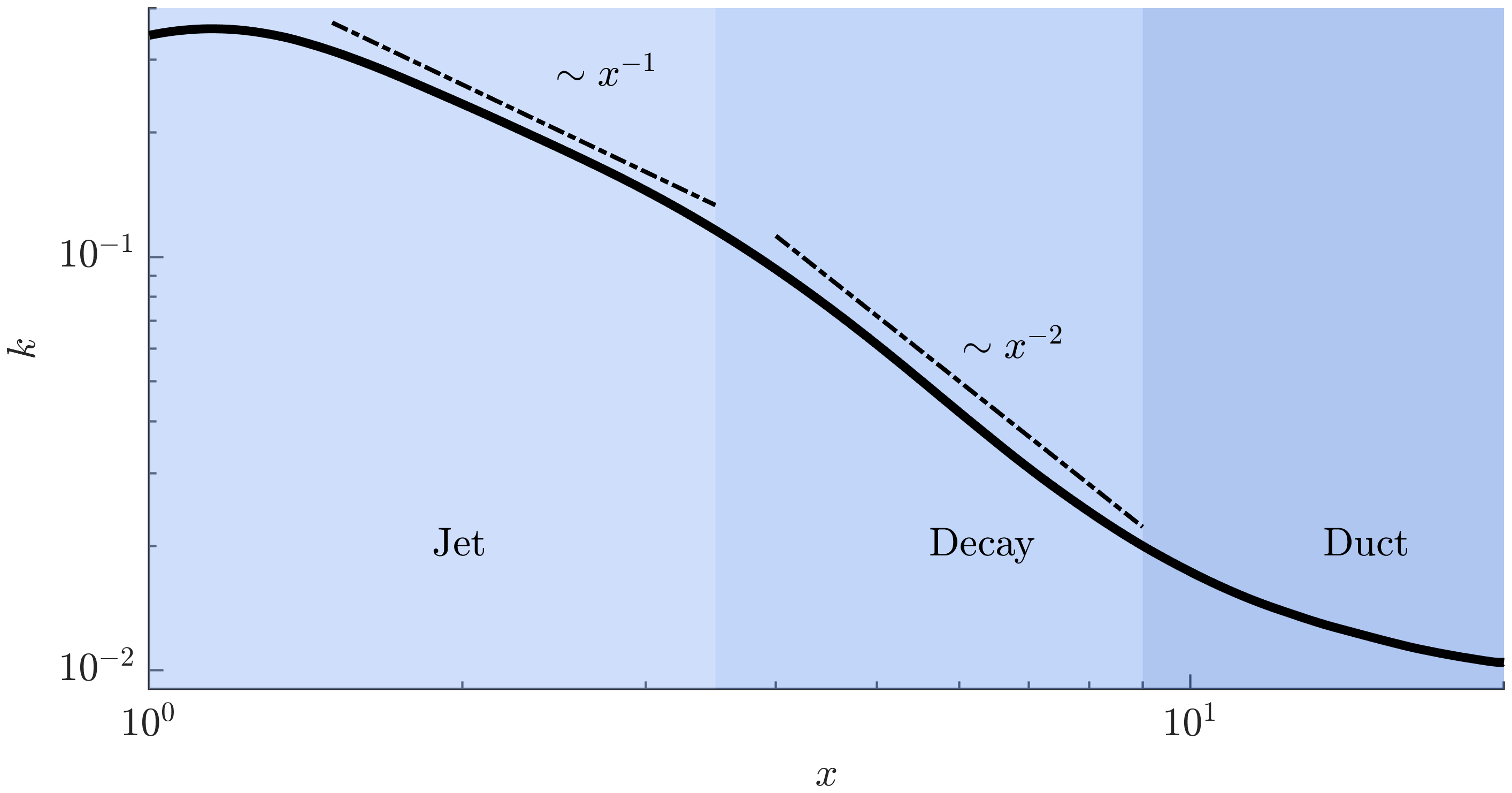}}
  \caption{Turbulent kinetic energy, $k$, averaged over the cross-section along the outlet at $Re=2000$. Different Flow regions are depicted by different shades of blue.}
\label{fig:regions}
\end{figure}

Both jet and decay regions exhibit power-law decreases of the kinetic energy (figure~\ref{fig:regions}). This further supports the analogies to a planar jet and homogeneous decaying turbulence in these regions, which will be studied in detail in \S\ref{sec:jet} and \S\ref{sec:decay}, respectively. The subsequent relaxation toward a duct flow is briefly discussed in 
\S\ref{sec:duct}.

\subsection{Jet-like flow}\label{sec:jet}

\begin{figure}
  \centerline{\includegraphics{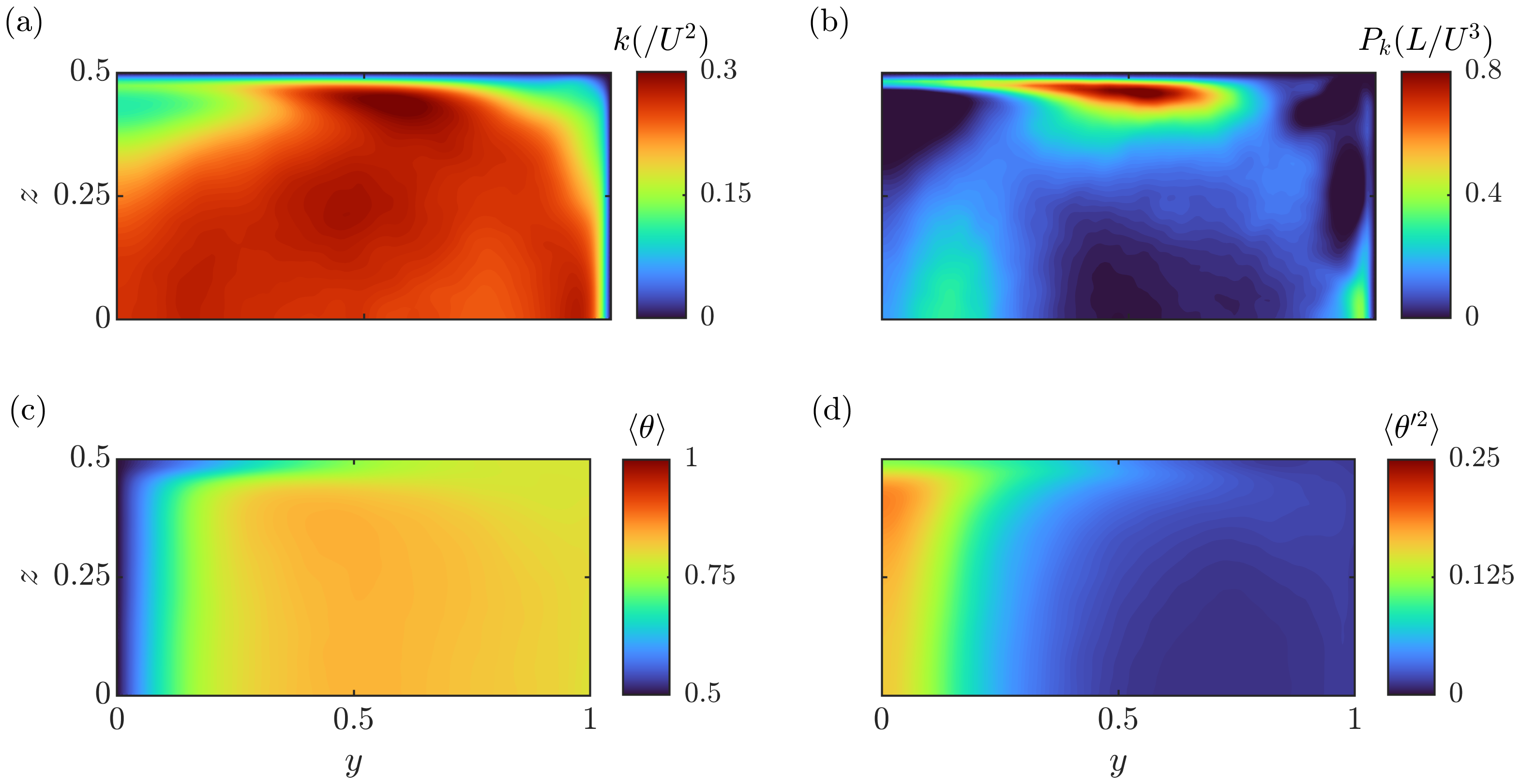}}
  \caption{ Cross-sectional colormaps of (a) $k$, (b) $P_{k}$, (c) $\langle\theta\rangle$ and (d) $\langle \theta^{\prime 2}\rangle $ for $Re=2000$ at $x=2$.}
\label{fig:cont2}
\end{figure}

In figure~\ref{fig:cont2}, we show colormaps of turbulent kinetic energy $k$, production $P_{k}$, 
and scalar average and variance at $x=2$. Owing to the left–right and up–down mirror symmetries of the mean turbulent flow, averages were taken over the four quadrants. The laminar inflowing streams are quasi-parabolic and thus carry most of the kinetic energy near the centreline. At the junction, where the flow transitions to turbulence, the vertical velocity profile flattens and as a consequence of the momentum redistribution strong shear layers with high values of  $k$ and  $P_{k}$ arise near the vertical walls. However, these do not contribute much to mixing (figure \ref{fig:cont2}(c)) because they are located away from the center. Additionally, elevated values of $P_k$ are observed near the center due to the collision of the inlet streams forming the jet. The resulting jet turbulence feeds the incipient mixing layer, and enhances mixing, as reflected in the scalar variance (figure \ref{fig:cont2}(d)).

\begin{figure}
  \centerline{\includegraphics{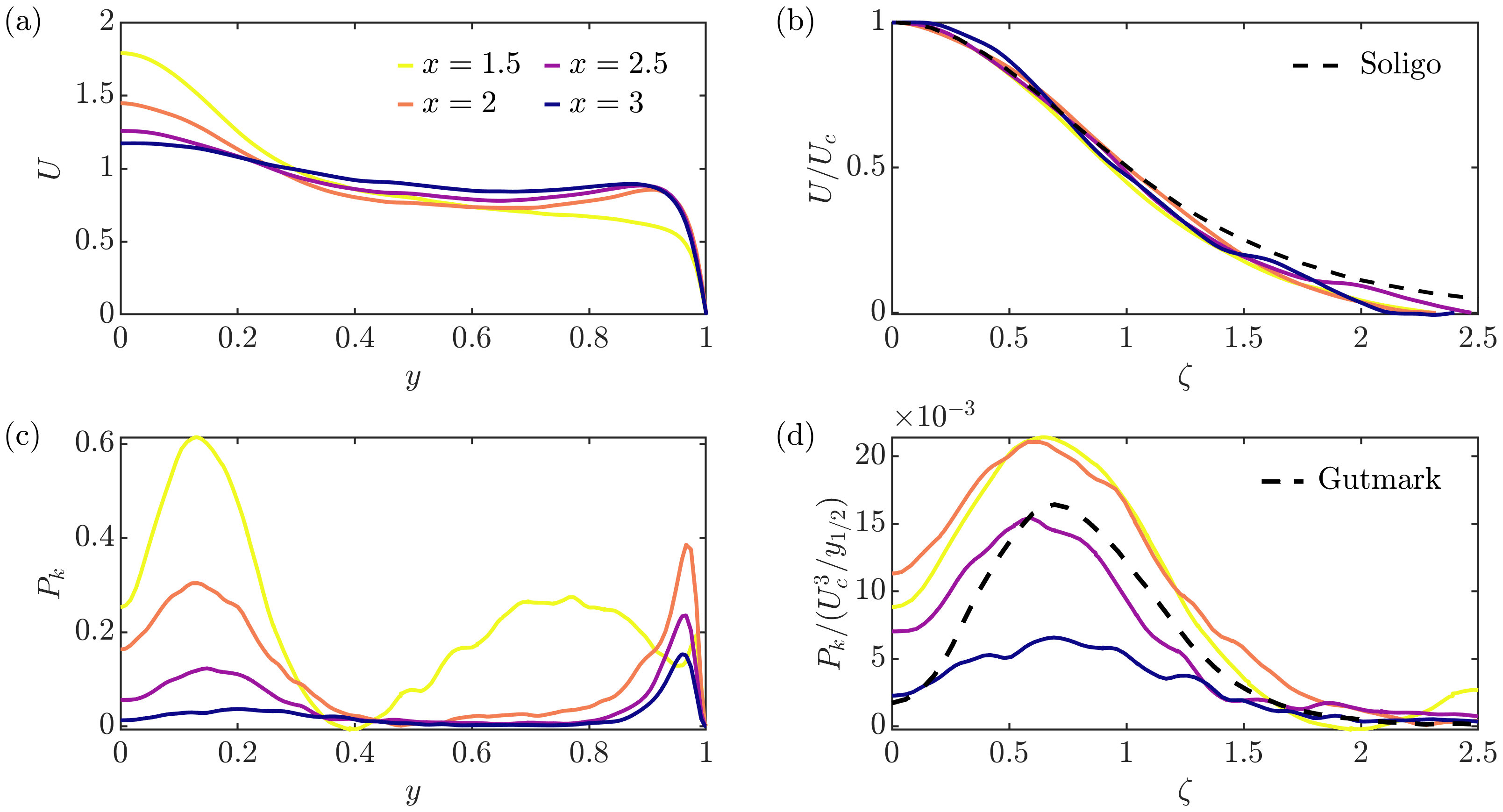}}
  \caption{(a) Velocity, (b) normalised velocity, (c) $P_{k}$ and (d) normalised $P_{k}$ profiles at $Re=2000$ and several positions along the jet region. The dashed line jet profiles in (b) and (d) are from \citet{Rosti_jet} and \citet{Gutmark}, respectively.}
\label{fig:jetfixed}
\end{figure}

In figure~\ref{fig:jetfixed}(a), stream-wise velocity profiles are shown at several locations along the jet region. To enable comparison with a planar jet, the mean velocity for $y \geq 0.5$ was first evaluated and subtracted from the velocity profiles. The jet half-width $y_{1/2}$ was then determined from these mean-subtracted profiles following the standard approach for co-flow jets \citep{coflow,coflow2}. In jet flows, velocity profiles are typically normalised by the centreline velocity ($U_{c}$) to asses self-similarity behaviour, while the transverse coordinate $y$ is scaled either by the jet half-width $y_{1/2}$, as $\zeta=y/y_{1/2}$, or by the axial distance from the jet origin $(x_{0})$, as $\eta=y/(x-x_{0})$. In planar jets, the spreading rate is linear, $\frac{dy_{1/2}}{dx}=S$, where $S\approx 0.1$ \citep{Pope_2000}. 

Using the resulting scaling $\eta=0.1\zeta$ within the jet region, the velocity profiles collapse reasonably well onto the theoretical planar jet profile up to $x\leq 3$ (figure \ref{fig:jetfixed}(b)). The normalised production $P_{k}$ profiles are shown in figure \ref{fig:jetfixed}(d), where the dashed line represents the benchmark data from \citet{Gutmark} for planar jet at $Re=3\times10^4$. The closest agreement with their data occurs at $x=2.5$. Nevertheless, it is evident that the self-similarity observed in the velocity profiles is weaker in the production curves. We attribute this discrepancy to the fact that planar jets at $Re=2000$ fall within the transitional regime. Specifically, \citet{Das_jet} reported that full self-similar behaviour in planar jet flows is first reached at $Re > 4000$. In the transitional regime, the flow is dominated by large coherent structures, and the mean velocity field attains self-similarity much earlier than the higher-order statistics \citep{Das_jet}. Accordingly, we observe a good collapse of the velocity profiles, but a weaker self-similar behaviour in the production profiles.
\begin{figure}
  \centerline{\includegraphics{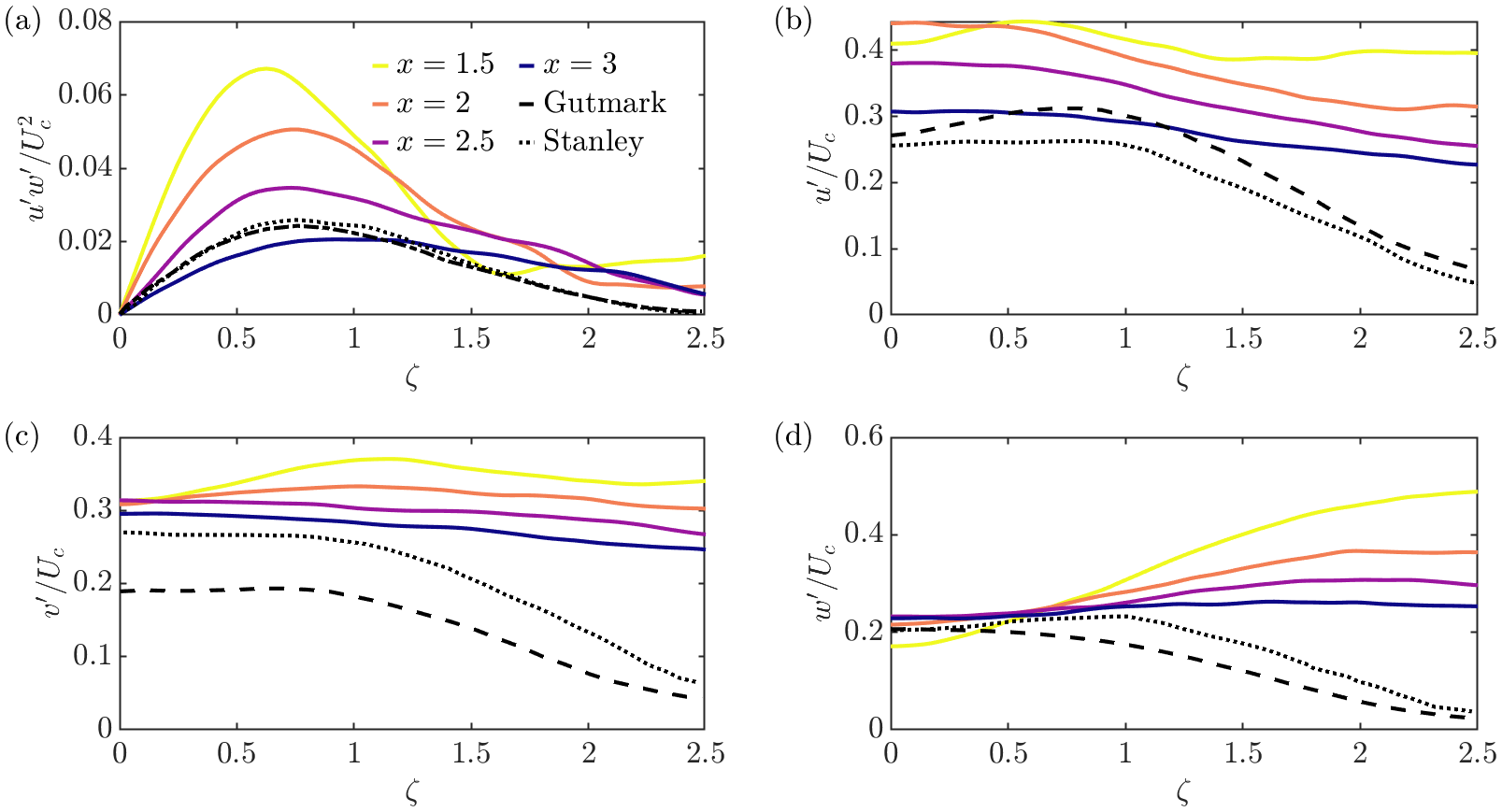}}
  \caption{(a) Reynolds shear stress (b) stream-wise (c) horizontal and, (d) vertical velocity fluctuations at $Re=2000$. The dotted and dashed lines are from \citet{stan} and \citet{Gutmark}, respectively.}
\label{fig:Urms}
\end{figure}

Reynolds shear stress and velocity fluctuation profiles are shown in figure~\ref{fig:Urms}(a)–(d) at four streamwise locations. Planar-jet measurements are included as reference \citep{Gutmark,stan}. In the central part of the outlet, the Reynolds shear stress and the fluctuations computed from our data exhibit similar profiles and magnitude as the planar-jet data. However, the agreement deteriorates toward the outer part of the cross-section. The difference is particularly pronounced in figure ~\ref{fig:Urms}(d). Here the fluctuation level in the T-mixer increases toward the outer region, whereas it decreases in the planar-jet measurements. This difference arises due to the fundamentally different flow configurations. A planar jet develops surrounded by a quiescent ambient fluid, leading to a progressive decay of the velocity fluctuations outside the shear layer. This decay in turbulence intensity is reflected in the turbulent production profile. The T-mixer outlet, by contrast, is a confined, wall-bounded flow in which the outer part of the cross-section is also turbulent. Furthermore, the finite distance between the developing shear layer and the sidewalls and the three-dimensional mean flow generated at the T-junction, modify the local strain field and the redistribution of turbulent kinetic energy among the velocity components. The planar-jet analogy therefore is limited to the free shear layer in the central region of the outlet channel. The deviations in the outer region arise due to confinement effects, the turbulent flow alongside the jet and three-dimensionality.
\begin{figure}
  \centerline{\includegraphics{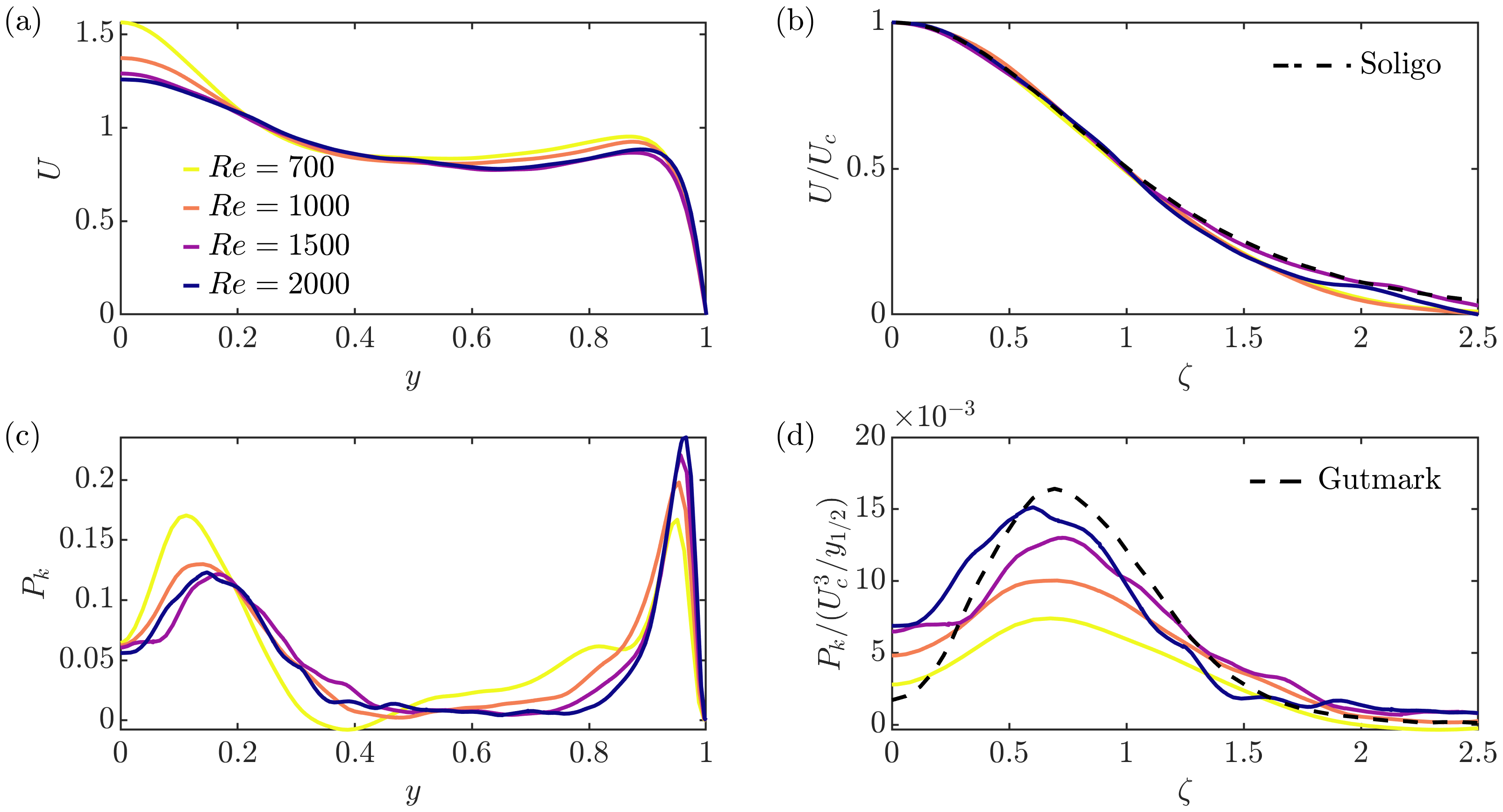}}
  \caption{Same as figure~\ref{fig:jetfixed} but at $x=2.5$ and several $Re$.}
\label{fig:jet}
\end{figure}

The velocity and production profiles for all $Re$ investigated in this work are plotted at $x=2.5$ in figure~\ref{fig:jet}. As expected, in all cases the velocity profiles exhibit the self-similar behaviour characteristic of planar jets, whereas the production profiles increasingly depart from the self-similar limit as $Re$ decreases.

\begin{figure}
  \centerline{\includegraphics{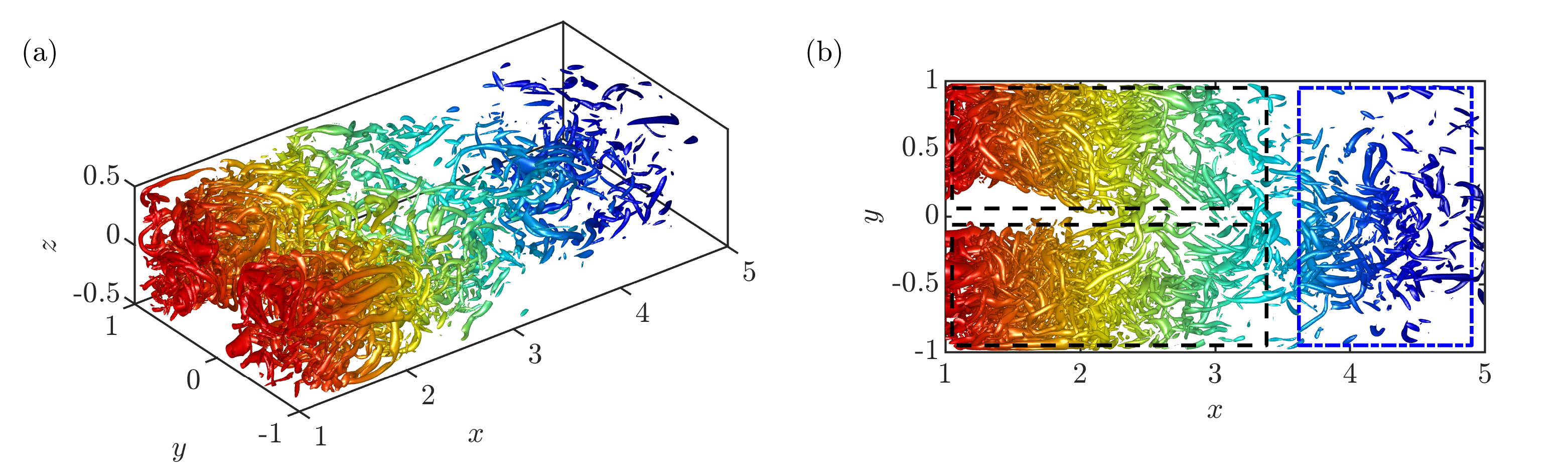}}
  \caption{Vortex structures defined by iso-surfaces of $Q$ criterion \citep{Q} at 95\% level in the jet region at $Re=2000$: (a) 3D view and (b) top view. The coloring is based on the stream-wise direction.}
\label{fig:Q1}
\end{figure}

The coherent vortex structures in the jet region are similar to those reported in studies of planar jets (figure~\ref{fig:Q1}(a)). Specifically, it can be observed that following the collision of the two inlet streams, pairs of vortical core structures are formed. These are highlighted in figure \ref{fig:Q1}(b) within the black dotted rectangles, similar to those reported for planar jets by \citet{Nan_jet} and \citet{Rosti_jet}. The vortices are initially oriented in the span-wise direction and are densely packed near the collision zone, where strong shear and turbulence production occur. As the flow progresses downstream, the two vortical cores elongate in the stream-wise direction. In the merging region, indicated in figure \ref{fig:Q1}(b) with a blue rectangle, the vortical pairs interact and combine into more organised coherent structures in the flow direction.

\subsection{Decay region}\label{sec:decay}

\begin{figure}
  \centerline{\includegraphics{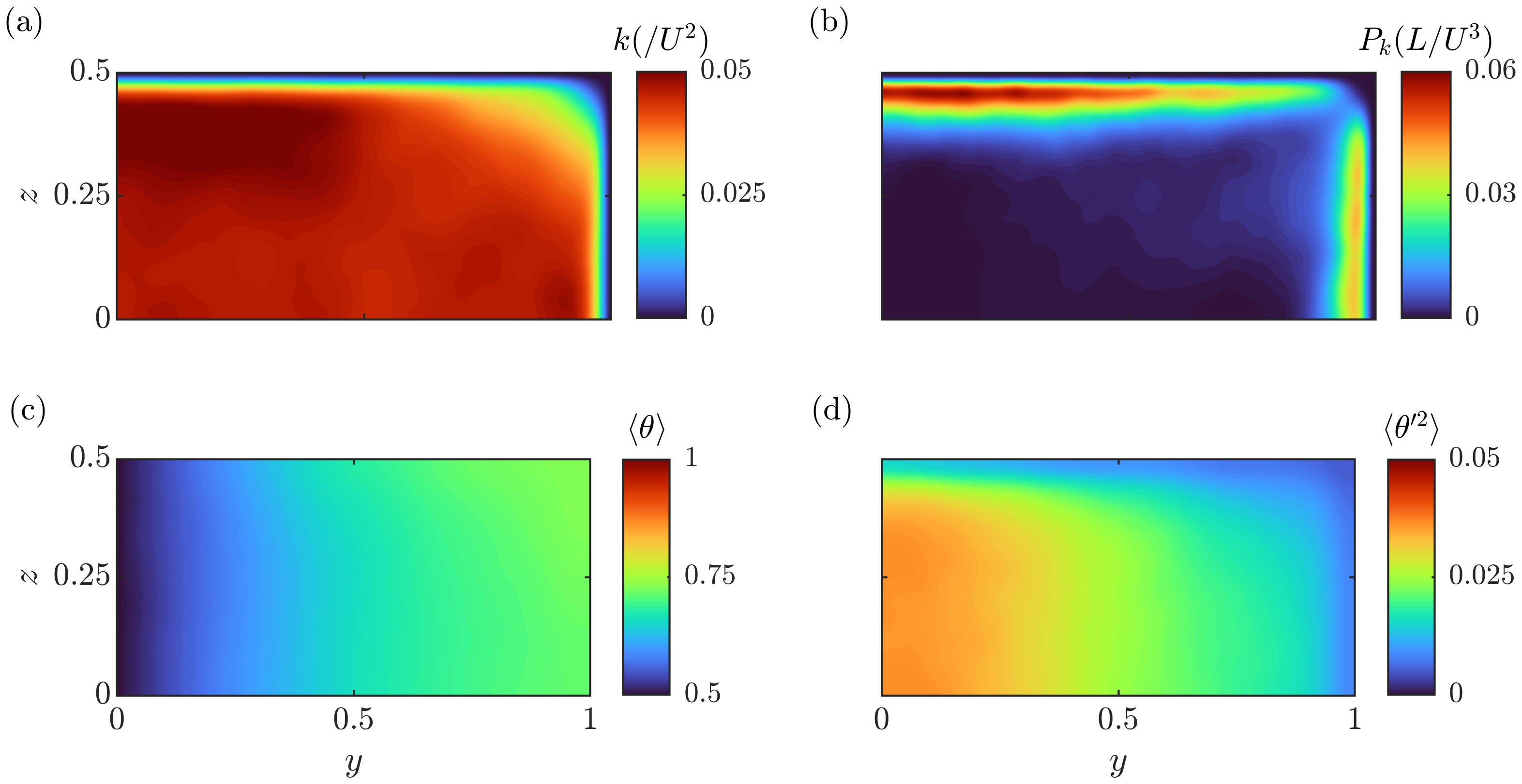}}
  \caption{As figure~\ref{fig:cont2} but at $x=6$.}
\label{fig:cont6}
\end{figure}

At $x=6$ the average stream-wise velocity profiles are flat (figure~\ref{fig:vel_profile}) and the kinetic energy has diffused across the entire cross-section (figure~\ref{fig:cont6}(a)), indicating that the jet has fully spread to the walls and the flow state is nearly homogeneously turbulent. The magnitude of $k$ here is an order of magnitude lower than at $x=2$, whereas its maximum value remains near the top and bottom walls. The production is also an order of magnitude lower and is strongly confined to near-wall layers (figure \ref{fig:cont6}(b)). The scalar field exhibits a distinctly different distribution. The averaged concentration features a smooth gradient in the span-wise direction from the bulk toward the walls, corresponding to a transition from fully mixed to largely unmixed inlet streams (figure \ref{fig:cont6}(c)). In the wall-normal direction, the mixing state is homogeneous. The scalar variance is roughly an order of magnitude lower than at $x=2$ and exhibits a similar span-wise distribution as the average, except for thin low-variance layers along the walls (figure~\ref{fig:cont6}(d)). 

The general picture remains nearly unchanged along the decay region, while the magnitudes of $k$, $P_k$, the span-wise gradient of $\langle\theta\rangle$ and $\langle \theta^{\prime 2} \rangle$ decrease monotonically.
\begin{figure}
  \centerline{\includegraphics{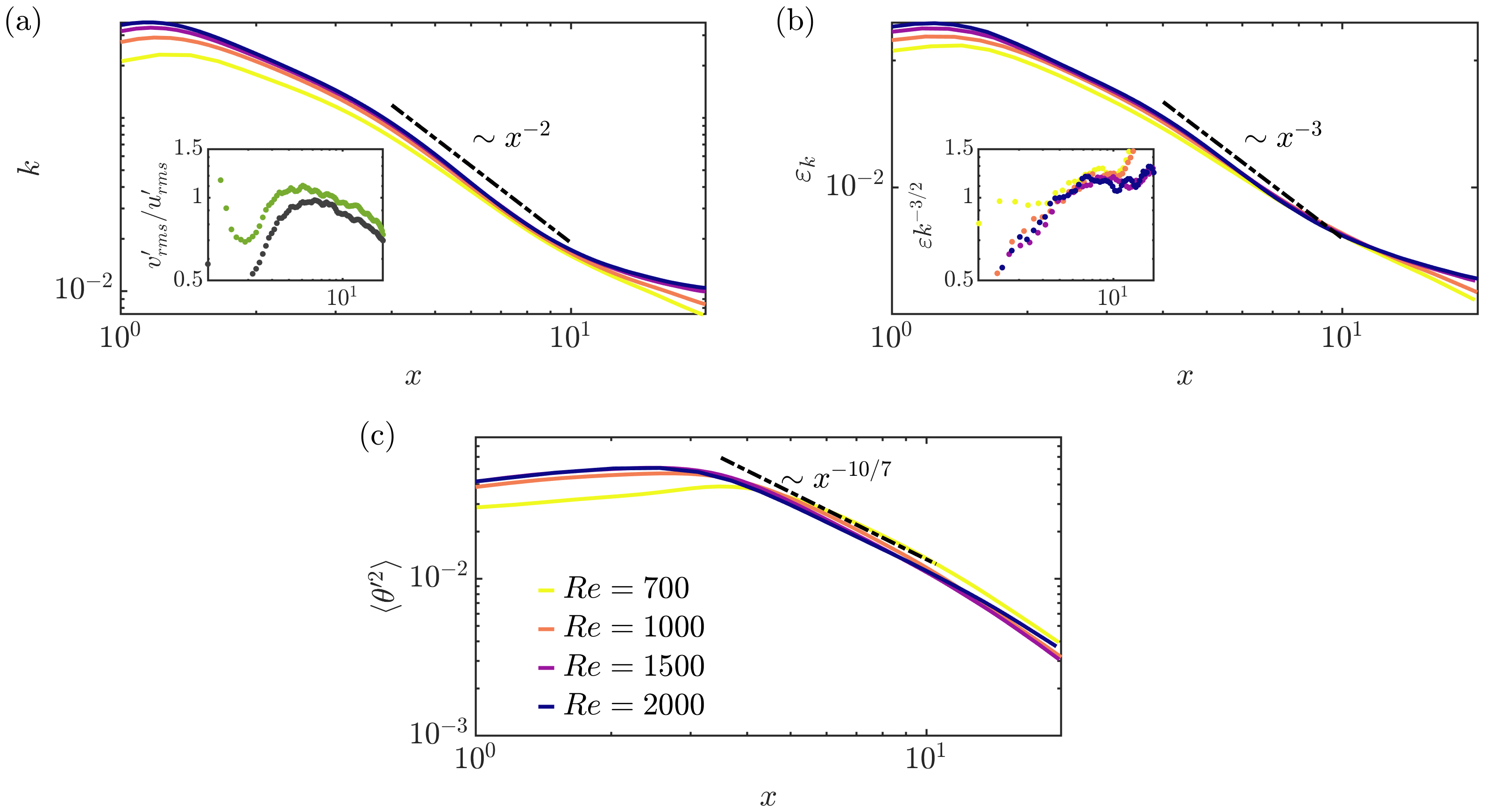}}
  \caption{(a) Evolution of the cross-sectionally averaged kinetic energy $k$, (b) dissipation $\varepsilon_{k}$ and (c) scalar variance $\theta^\prime$ along the outlet at various $Re$. The inset in (a) shows $v'_\text{rms}/u'_\text{rms}$ (green) and $w'_\text{rms}/u'_\text{rms}$ (black) along the centreline for $Re=2000$. The inset in (b) shows $c_{\varepsilon}=\varepsilon \, k^{-3/2}$ along the centreline.}
\label{fig:full_decay}
\end{figure}
Specifically, we show in figure~\ref{fig:full_decay}(a) that the cross-sectionally averaged turbulent kinetic energy decreases as a power-law with exponent $-2$ at all $Re$ investigated. As shown in the inset, the flow conditions are nearly isotropic near the centreline, where $v'_\text{rms}/u'_\text{rms} \approx w'_\text{rms}/u'_\text{rms} \approx 1$. The dissipation rate also exhibits a power-law decay, but with an exponent of $-3$ (figure~\ref{fig:full_decay}(b)). As shown in the inset of figure \ref{fig:full_decay} (b), $\varepsilon \ k^{-3/2}$ becomes approximately constant in the decay region and approaches unity as $Re$ increases, indicating the persistence of fully developed turbulent flow in the decay region. 
In the (approximately) isotropic region of the outlet, kinetic energy expected to decay as
\begin{equation}\label{k}
\frac{dk}{dx}=-\varepsilon,
\end{equation}
where the dissipation rate is modeled as $\varepsilon = c k^{3/2}/l$, and $l$ is the integral length scale. In a confined flow, $l$ is the size of the domain. Substituting this estimate into~\eqref{k} and solving for $k$ yields a power-law decay of the form $k(x)\sim x^{-2}$ and $\varepsilon(x)\sim x^{-3}$, which is consistent with the decay exponents observed in our DNS.
Additionally, the observed decay exponents agree with the values reported by \citet{bounded1} and \citet{bounded2} for turbulence decay in bounded domains which are significantly higher in magnitude than those reported for homogeneous isotropic turbulence \citep{unbounded1}. This is consistent with the fact that in the decay region of the T-mixer production is confined to the near-wall layers.

The evolution of the scalar variance along the outlet is shown in figure \ref{fig:full_decay}(c). At the junction, the two streams are initially largely unmixed and begin to mix in the jet region, leading to an increase in the scalar variance. Meanwhile, the turbulence intensity increases and contributes to decreasing the variance. This leads to a maximum in the variance, which is achieved earlier, the more turbulent the flow is (i.e. at higher $Re$). Subsequently, the variance decays monotonically. In the decay region, it eventually follows a power-law decay with an exponent of approximately $-10/7$. This is consistent with the scalar variance decay reported for homogeneous isotropic turbulence by \citet{chasnov} and \citet{Scalar1}, and clearly different from the exponent for decay in bounded domains ($-3.25$) reported by \citet{Scalar2}. We attribute this to the fact that mixing is initially localized in the vertical mid-plane of the channel and slowly progresses toward the side walls (compare figures~\ref{fig:cont2}(d) and \ref{fig:cont6}(d)). Even at $x=6$, the scalar variance remains confined to the central region, where the flow is approximately homogeneous and isotropic. 
\subsection{Relaxation to duct flow}\label{sec:duct}
\begin{figure}
  \centerline{\includegraphics{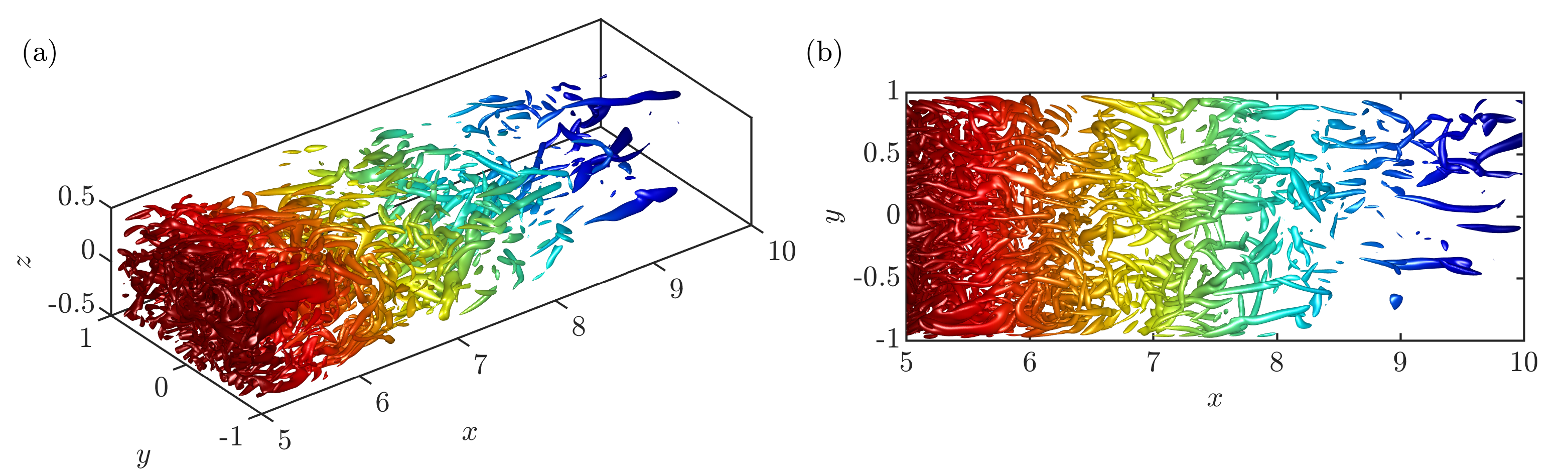}}
  \caption{As in figure~\ref{fig:Q1} for the decay region.}
\label{fig:Q2}
\end{figure}
\begin{figure}
  \centerline{\includegraphics{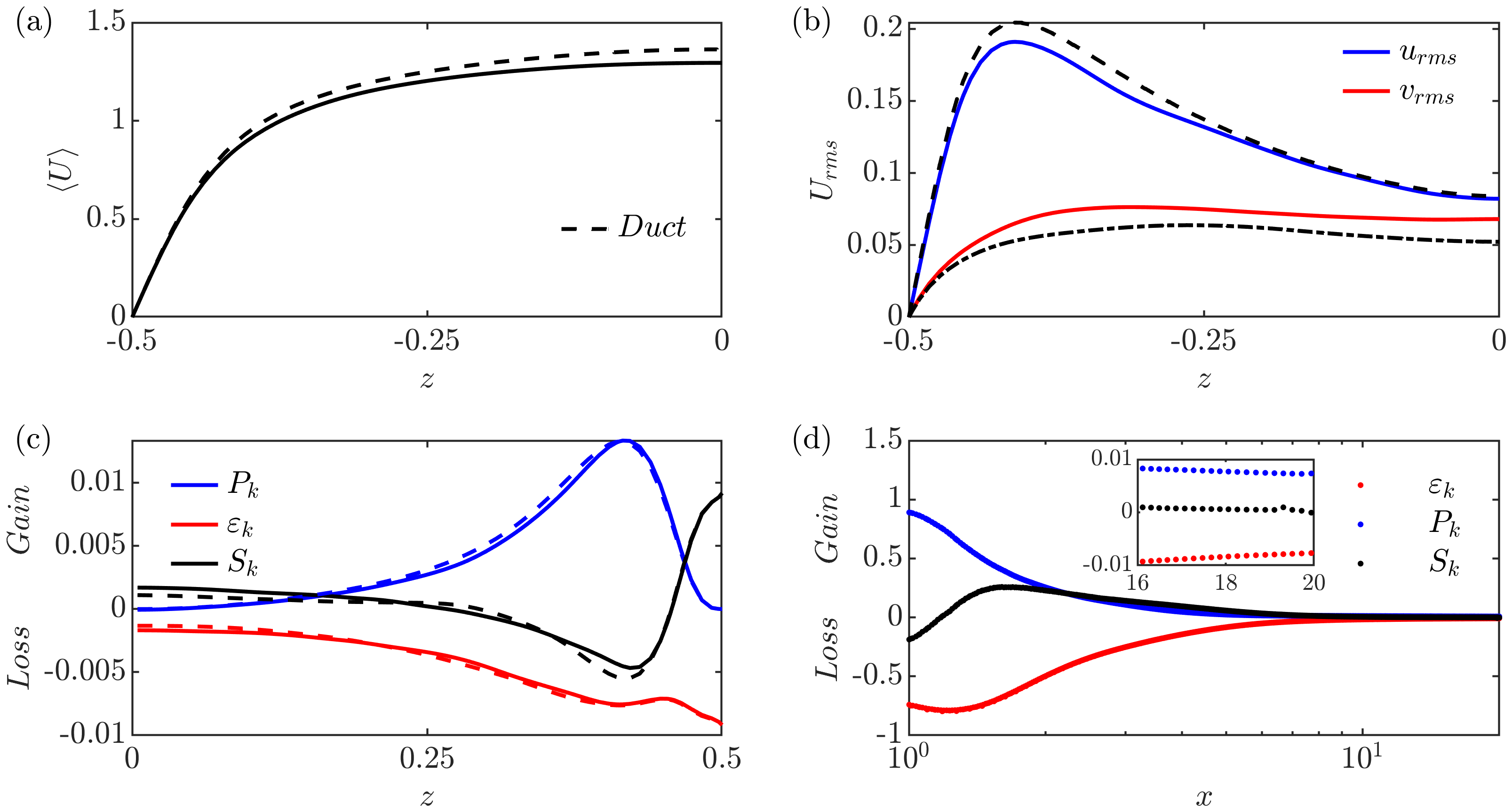}}
  \caption{Stream-wise velocity profile (a), root-mean-square velocities (b) and production $P_{k}$, $\varepsilon_{k}$ and sum of TKE-fluxes $S_{k}$ along the wall-normal direction ($y=0$ and $x=18$) (c). The dashed line represents the rectangular duct data with aspect ratio of $2$ at $Re=2000$. (d) Cross-sectionally integrated $P_{k}$, $\varepsilon_{k}$ and $S_{k}$ along the outlet (d).}
\label{fig:rms_final}
\end{figure}

Along the decay region, the turbulent structures become progressively less intense and grow in size (figure~\ref{fig:Q2}). From approximately $x\gtrsim9$ onward, they begin to elongate and align in the stream-wise direction, marking the beginning of the gradual relaxation toward duct flow. This increase in the anisotropies is quantified in the inset of figure~\ref{fig:full_decay}(a)). The kinetic energy and dissipation rate begin to level off and relax toward turbulent duct-flow values for $Re=1500$ and $2000$, whereas they continue to fall sharply for $Re=700$ and $1000$ (figure \ref{fig:full_decay}(a)--(b)), indicating flow relaminarisation. Note also how for the latter cases $\varepsilon k^{-3/2}$ increases linearly as a result of the dissipation falling off less sharply than the kinetic energy during the relaminarisation process, whereas for the former it begins to saturates at about $\varepsilon k^{-3/2} \rightarrow 1.25$, as in turbulent duct flow, with $\varepsilon k^{-3/2} \approx 1.4$  from our computations. However, we stress that the outlet length simulated here ($20$) does not allow for full development of the duct flow. Specifically, at $x=18$ and $z=0$ the mean velocity profile is flatter than in duct flow (figures~\ref{fig:rms_final}(a)), and the turbulent intensity is higher (figures~\ref{fig:rms_final}(b)). 
\
\\
In figure \ref{fig:rms_final}(c), the TKE budget terms at $x=18$ and $y=0$ are examined. Production $P_{k}$ and dissipation $\varepsilon_{k}$ are confined to the near wall regions, and there is a strong flux of kinetic energy, denoted by $S_k$, from the wall toward the bulk. The dashed curves correspond to the reference duct data and indicate that, while the flow is converging and relaxing toward duct flow condition, residual turbulence persists and a fully developed duct flow state has not yet been attained. Finally, we shown in figure \ref{fig:rms_final}(d) the cross-sectionally integrated TKE budget terms along the outlet. Throughout the jet region there is a strong down-stream flux of kinetic energy, which weakens but persists throughout the decay and relaxation regions.
\\
\begin{center}
\centering
  \includegraphics{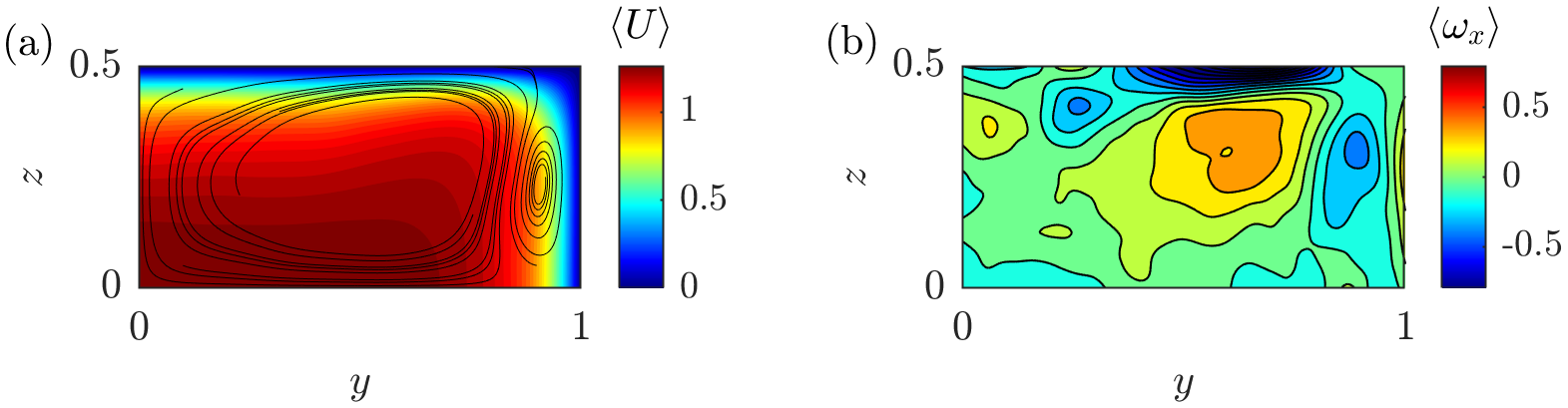}
  \captionof{figure}{(a) Cross-sectional colormap of mean velocity magnitude and streamlines at $x=18$. (b) Colormap of mean stream-wise vorticity at $x=18$.}
\label{fig:contduct}
\end{center}
\
\\
Streamlines of the mean velocity field and stream-wise vorticity are shown in figure~\ref{fig:contduct}. The two-vortex pattern in the corner, which is typical of turbulent duct flows, is observed here. The vortex adjacent to the horizontal wall is larger and elongated toward the center, while the smaller vortex remains closer to the vertical wall. This is as expected for ducts with an aspect ratio of 2, which break the symmetry between the two corner vortices about the corner bisector characteristic of square ducts.

\subsection{Mixing process}
\begin{figure}
  \centerline{\includegraphics{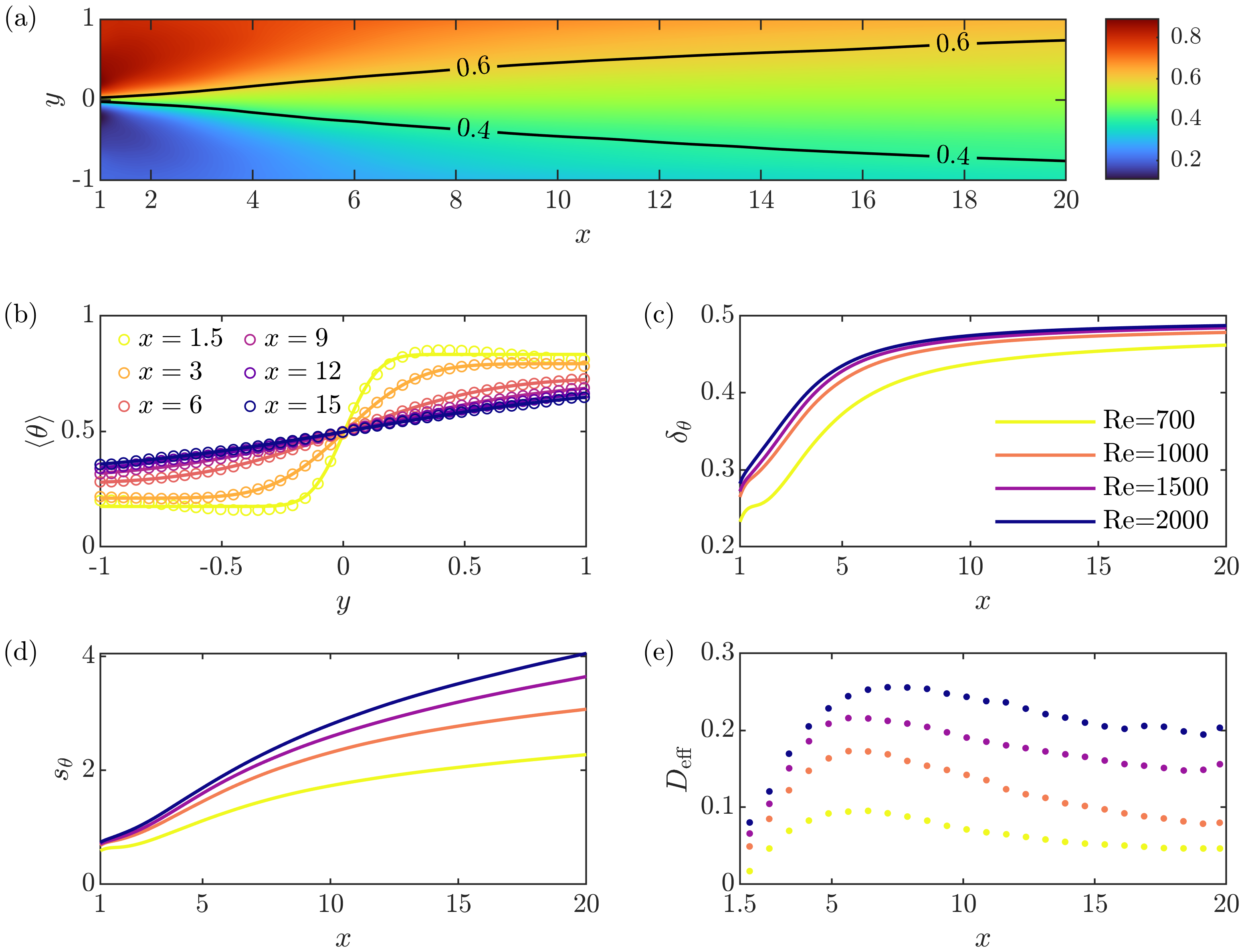}}
  \caption{(a) Span-wise colormap of mean scalar $\langle \theta \rangle$ averaged in vertical direction ($z$). (b) Scalar profiles at various streamwise locations, with the corresponding error-function fits shown as solid lines. (c) Computed $\delta_{\theta}$ from~\eqref{deltaeq} for various $Re$. (d) Mixing thickness computed using ~\eqref{integ} for various $Re$.(e) Computed $D_{\mathrm{eff}}$ using~\eqref{Deffq} along the outlet.}
\label{fig:mixingf}
\end{figure}

\begin{figure}
  \centerline{\includegraphics{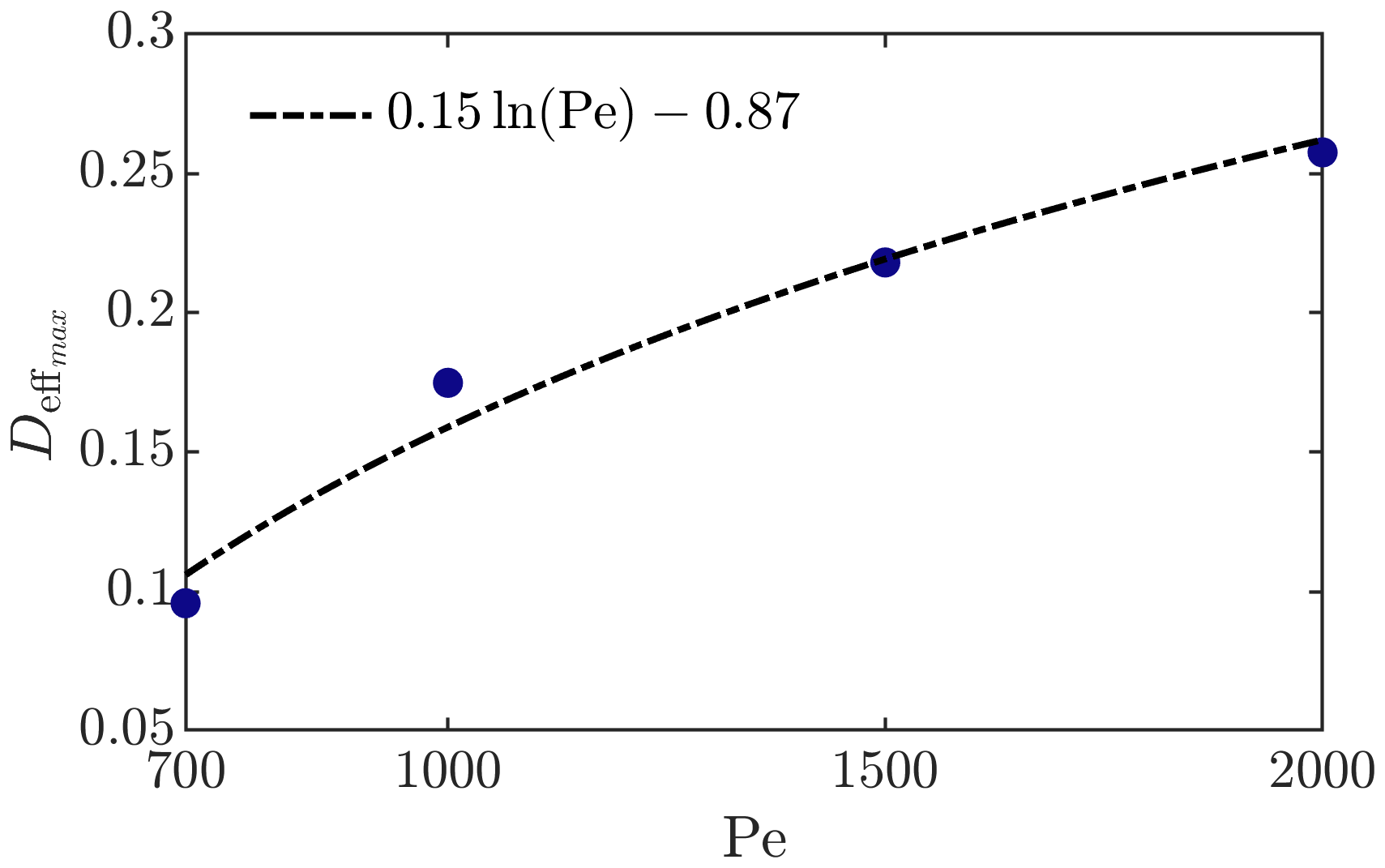}}
  \caption{Scaling of the maximum effective diffusion coefficient with respect to Pe number.}
\label{fig:scalingPe}
\end{figure}

Figure~\ref{fig:mixingf}(a) shows a colormap of the temporally and vertically averaged scalar field, $\langle \theta \rangle$, along the outlet. The thin jet sheet created at the junction expands toward the walls as the two streams gradually mix. The two black lines denote contour levels at 0.4 and 0.6, and are plotted to highlight the expansion of the mixing boundaries.
As shown in the Appendix~\ref{Appendix_A}, the following advection-diffusion equation  approximately describes the mixing process in the main channel of the T-mixer
\begin{equation}\label{adv2}
    U\frac{\partial\langle \theta \rangle}{\partial x}=D_{\mathrm{eff}}\frac{\partial^2\langle \theta \rangle}{\partial y^2}
\end{equation}
Here $D_{\mathrm{eff}}=D+D_{t}$ is the effective diffusion coefficient and $D_{t}$ is a turbulent eddy diffusivity. For a Heaviside initial condition of the form
\begin{equation}
    \langle \theta(x=x_{0},y) \rangle= 
    \begin{cases}
0  & y < 0 \\
1  & y \ge 0,
\end{cases}
\end{equation}
the solution to~\eqref{adv2} is an error function of the form~\citep{crank}
\begin{equation}\label{erf}
    \langle \theta(x,y) \rangle=\frac{1}{2}[1+\mathrm{erf} (\frac{y}{s_{\theta}})]
\end{equation}
where $s_{\theta}=\sqrt{4xD_{\mathrm{eff}}/U}$ is the slope of the error function profile and represents the streamwise evolution of the mixing thickness. Although the scalar profiles are not generated from a true Heaviside initial condition in the outlet, they resemble an error-function shape (see figure~\ref{fig:mixingf}(b)). Therefore, an error-function fit to the mean scalar profile can be used to estimate the effective diffusion coefficient, $D_{\mathrm{eff}}$, along the outlet of the T-mixer. This, in turn, allows us to quantify how each flow region contributes to the overall mixing process. However, this approach has two limitations. First, one must choose the effective origin location $x_{0}$ used to evaluate $x$ in the expression for $s_{\theta}$. Second, the fitting is sensitive to noise, and the resulting estimates can be quite noisy, masking the underlying physical trend of $D_{\mathrm{eff}}$ along the outlet. An alternative approach is to use scalar integral mixing width $\delta_{\theta}$ to compute the effective diffusion coefficient, since it is an integral quantity and therefore less sensitive to noise.
For a freely evolving mixing layer, the scalar integral mixing width is expressed as \citep{Blakeley}
\begin{equation}\label{deltaeq}
    {\delta}_{\theta}=\int_{-\infty}^{\infty}\langle \theta \rangle(1-\langle \theta \rangle)dy.
\end{equation}
In the T-mixer the domain is bounded, so the integration limits in the above expression must be replaced by the duct lateral boundaries, i.e., from $y=-1$ to $y=1$. For all cases, the integral mixing width $\delta_{\theta}$ increases significantly in the jet region and then exhibits a saturating trend toward values around $0.5$ at the decay and duct region (see figure~\ref{fig:mixingf}(c)). This is expected, since $0.5$ is the maximum value $\delta_{\theta}$ can attain based on its definition. These values of $\delta_{\theta}$ for each case can then be used to compute the effective diffusion coefficient along the outlet. To compute $D_{\mathrm{eff}}$ from $\delta_{\theta}$, we first need to establish a relation between the two.
Substituting~\eqref{erf} into the definition of the integral mixing width,~\eqref{deltaeq}, and integrating over the spanwise direction yields (see Appendix~\ref{Appendix_B})
\begin{equation}\label{integ}
    \delta_{\theta}=\frac{1}{2}-\frac{1}{2}\mathrm{erf}^2(\frac{1}{s_{\theta}})-\frac{s_{\theta}}{\sqrt{\pi}}\mathrm{erf}(\frac{1}{s_{\theta}})e^{(-1/s_{\theta}^2)}+\frac{s_{\theta}}{2}\sqrt{\frac{2}{\pi}}\mathrm{erf}(\frac{\sqrt{2}}{s_{\theta}})
\end{equation}
 Equation~\eqref{integ} provides a direct relation between $\delta_{\theta}$ and $s_{\theta}$. Using~\eqref{integ}, the mixing thickness $s_{\theta}$ along the outlet can be computed and is plotted in figure~\ref{fig:mixingf}(d). Since the error function is a self-similar solution of Eq.~\eqref{adv2}, $s_{\theta}$ evolves downstream according to 
 \begin{equation}
     s_{\theta}(x+\Delta x)=\sqrt{s_{\theta}^2(x)+4\Delta xD_{\mathrm{eff}}/U}
 \end{equation}
 where $\Delta x$ is the distance between two stream-wise locations and $D_{\mathrm{eff}}$ the effective diffusion coefficient for the mixing process between these two locations. This resolves the problem of setting the origin, $x_{0}$, of the Heaviside function. Assuming $U=1$ at each cross-section, $D_{\mathrm{eff}}$ can be expressed as
\begin{equation}\label{Deffq}
D_{\mathrm{eff}}=\frac{s_{\theta}^2(x+\Delta x)-s_{\theta}^2(x)}{4\Delta x}.
\end{equation}
This expression relates $D_{\mathrm{eff}}$ to $\delta_{\theta}$ through the mixing thickness $s_{\theta}$; first, $s_{\theta}(x)$ is extracted from $\delta_{\theta}(x)$ using the~\eqref{integ}, and then the resulting $s_{\theta}(x)$ is used in~\eqref{Deffq} to compute $D_{\mathrm{eff}}$ along the outlet. 
\\
Figure~\ref{fig:mixingf}(e) shows the streamwise evolution of the effective diffusion coefficient, $D_{\mathrm{eff}}$, along the outlet of the T-mixer for different Reynolds numbers calculated with $\Delta x=0.5$. For all cases, $D_{\mathrm{eff}}$ increases sharply immediately downstream of the junction, reflecting the strong enhancement of scalar transport caused by the impingement of the two inlet streams and the formation of the jet region. In this region, intense interface stretching, entrainment, and transverse transport produce highly efficient mixing, leading to a rapid increase in the effective diffusivity. The effective diffusion coefficient reaches a maximum near the end of the jet region, indicating the location of strongest mixing activity. Further downstream, $D_{\mathrm{eff}}$ gradually decreases as the influence of the jet weakens, coherent structures decay, and scalar gradients become progressively smoother. Nevertheless, the effective diffusivity does not decay to zero, but appears to approach a finite asymptotic value for the two largest Reynolds number cases. This suggests that, for a sufficiently long outlet, these cases would likely approach an asymptotic effective diffusivity associated with fully developed duct flow dispersion at that specific $Re$. The relatively weak variation in the peak location with $Re$ suggests that the streamwise position of maximum mixing is governed primarily by the T-mixer geometry, while the Reynolds number mainly controls the intensity of the mixing process. Figure~\ref{fig:scalingPe} shows the scaling of the peak value of $D_{\mathrm{eff}}$ with the Péclet number. Within the investigated range of  700$\leq$ Pe$\leq$2000, the data are consistent
with a logarithmic trend, although the present dataset is insufficient
to distinguish this behaviour from other slowly varying functional forms.
Investigating a broader range of Péclet numbers, particularly higher Pe,
would be necessary to establish the underlying scaling and determine the
asymptotic trend. This apparent logarithmic trend, however, contrasts with classical laminar dispersion, where \citet{taylor} reported $D_{\mathrm{eff}} \sim \mathrm{Pe}^2$ and suggests that, in the T-mixer, increasing $Re$ or equivalently $\mathrm{Pe}$, does not significantly enhance mixing~\citep{Tobias2019}. 

\section{Conclusion}

Turbulence in the outlet of the T-mixer exhibits two distinct regimes analogous to two well-known canonical flows, followed by a slow relaxation toward duct flow. After the collision of the incoming streams at the junction ($1.5 \lesssim x \lesssim 3$), a high-velocity jet forms and displays the typical coherent vortical structures and statistics of a transitional planar jet. This includes a linear decrease of the kinetic energy  \citep{Pope_2000}, a very good agreement of the stream-wise velocity profiles with those of a self-similar planar jet, but less so for the production profiles \citep{Das_jet}. The velocity fluctuations and Reynolds shear stress show good agreement with previous studies \citep{Gutmark,stan} in the central region, but this agreement deteriorates near the walls as the jet widens and approaches the boundaries. Subsequently ($4\lesssim x \lesssim 9$), a fully developed, nearly homogeneous turbulent state with thin boundary layers at the walls emerges. It is characterised by a power-law decay of kinetic energy and dissipation rate with the exponents typical of decaying turbulence in bounded domains \citep{bounded1,bounded2}, i.e.\ $-2$ and $-3$ respectively.   By contrast, the decay exponent obtained for the scalar variance ($-10/7$) is in agreement with the values reported for unbounded turbulence \cite{Scalar1}. We interpret this apparently contradictory behaviour to the fact that the mixing statistics are nearly homogeneous in the wall-normal direction and slowly develop in the span-wise direction from the mid-plane towards the side walls without feeling their effect in the decay region.  For the laminar inlet boundary conditions employed here, the Reynolds number based on the Taylor microscale reaches a maximum of $Re_{\lambda}=\sqrt{10k^2/(\varepsilon\nu)}\sim 200$ at the centreline in the junction and $Re=2000$. It decreases down to $Re_{\lambda}\sim 26$ at the end of the outlet channel, indicating a relaxation toward a fully developed duct flow ($Re_{\lambda}\sim 24$), even if this cannot be achieved with the outlet length used here ($20d$).

The need for studying turbulence decay beyond the traditionally considered canonical configurations has been recently emphasised by \citet{Lohse} and \citet{singh2021turbulence}. Here, we demonstrated that the T-mixer provides a suitable configuration for investigating turbulence decay and jet flows that is accessible in both experiments and simulations, with well-defined reproducible boundary conditions. Although the stream-wise profiles of kinetic energy and dissipation are nearly identical at $Re=2000$ and $3000$ \citep{Tobias2019}, indicating saturation to the asymptotic turbulent state, investigating higher $Re$ than done here would enable probing the convergence toward a fully developed planar jet and the apparent logarithmic evolution of the effective diffusion with the Péclet number. 
\\
Simulations with turbulent inlets merit a separate investigation, as they exhibit a markedly different decay of the kinetic energy and dissipation \citep{Tobias2019} after the junction. We argue that this may result from the qualitatively different jet flow that must form when two turbulent (blunted) profiles collide at the junction. Finally, the T-mixer offers the possibility of studying active-scalar mixing in a well-defined, simple setting, which has not been attempted so far despite the relevance of active mixing in chemical engineering \citep{schikarski2023impact}. In active-scalar mixing, the scalar affects the velocity field. Therefore, a detailed understanding of the underlying hydrodynamics is necessary in order to separate the influence of the active scalar from the flow dynamics. The results presented here characterize the purely hydrodynamic behavior of the T-mixer in detail, and thus provide a solid baseline for future studies of active-scalar mixing in T-shaped mixers.



\backsection[Acknowledgements]{The authors would like to thank Dr. Daniel Morón Montesdeoca, Dr. Ianto Canoon, Patrick Keuchel and Felix Kranz for their insightful discussions during the revision process.}

\backsection[Funding]{This work was supported by the Deutsche Forschungsgemeinschaft (DFG, German Science Foundation) with Grant No. 511099203.}

\backsection[Declaration of interests]{The authors report no conflict of interest.}

\backsection[Data availability statement]{The data that support the findings of this study will be made openly available in \href{https://www.pangaea.de/}{Pangaea} at at http://doi.org/[doi].}



\appendix

\section{Approximated advection diffusion equation in the outlet}\label{Appendix_A}
We consider the advcetion diffusion equation 
\begin{equation}
    \frac{\partial\theta}{\partial t}+\textbf{u}\cdot\nabla\theta=D\Delta\theta
\end{equation}
and using the Reynolds decomposition, followed by averaging in time, one obtain
\begin{equation}\label{A1}
    \langle \mathbf{u} \rangle \cdot \nabla \langle \theta \rangle
+
\left\langle \mathbf{u}' \cdot \nabla \theta' \right\rangle
=
D \nabla^2 \langle \theta \rangle.
\end{equation}
where $\langle \textbf{u}\rangle=(\langle u\rangle,\langle v\rangle,\langle w\rangle)$ and $\textbf{u}'=(u',v',w')$ denote the mean and fluctuating velocity fields. The~\eqref{A1} can be rearranged as
\begin{equation}\label{A3}
\langle \mathbf{u} \rangle \cdot \nabla \langle \theta \rangle
=
D \nabla^2 \langle \theta \rangle
-
\nabla \cdot \langle \mathbf{u}' \theta' \rangle .
\end{equation}
where $\langle \textbf{u}'\theta'\rangle=(\langle u'\theta'\rangle,\langle v'\theta'\rangle,\langle w'\theta'\rangle)$ is the turbulent scalar flux vector.We then average the fields in the vertical direction, $z$, which yields $\langle w\rangle=0$. Additionally, since the main flow in the outlet is in the streamwise direction $x$, we can assume $\langle v\rangle \approx0$. It can also be seen in figure~\ref{fig:mixingf}(a) that the scalar gradients are much larger in the spanwise direction $y$ than the stream-wise $x$ direction, and the mixing process is mainly a transverse mixing in the $y$ direction. Accordingly~\ref{A3} reduces to
\begin{equation}\label{A4}
     \langle u \rangle\frac{\partial \langle \theta \rangle}{\partial x}=D\frac{\partial^2\langle \theta\rangle}{\partial y^2}-\frac{\partial \langle v'\theta' \rangle}{\partial y}.
\end{equation}
We use the first order eddy diffusivity model \citep{Pope_2000}  
\begin{equation}
    \langle v'\theta'\rangle=-D_{t}\frac{\partial \langle \theta\rangle}{\partial y},
\end{equation}
and assume that $D_{t}$ does not depend on $y$, and obtain
\begin{equation}
    \langle u \rangle\frac{\partial \langle \theta \rangle}{\partial x}=(D+D_{t})\frac{\partial^2\langle \theta\rangle}{\partial y^2}.
\end{equation}
\section{Integral mixing width expression}\label{Appendix_B}
Starting from the error function profile~\eqref{erf} and inserting it into the definition of integral mixing width~\eqref{deltaeq} we get
\begin{equation}\label{B1}
    \delta_{\theta}=\frac{1}{2}-\frac{s_{\theta}}{4}\int_{-1/s_{\theta}}^{1/s_{\theta}}\mathrm{erf^2(\lambda)} d\lambda
\end{equation}
where $\lambda=y/s_{\theta}$. Since the $\mathrm{erf}^2(\lambda)$ is even, we write the integral term as \citep{erfbook}
\begin{equation}\label{B2}
    I=2\int_{0}^{1/s_{\theta}}\mathrm{erf^2(\lambda)}d\lambda=2\lambda\mathrm{erf^2}(\lambda)+\frac{4}{\pi}\mathrm{erf}(\lambda)e^{-\lambda^2}-\frac{2\sqrt{2}}{\sqrt{\pi}}\mathrm{erf}(\sqrt{2}\lambda).
\end{equation}
Inserting~\eqref{B2} into~\eqref{B1}, we get 
\begin{equation}\label{B4}
        \delta_{\theta}=\frac{1}{2}-\frac{1}{2}\mathrm{erf}^2(\frac{1}{s_{\theta}})-\frac{s_{\theta}}{\sqrt{\pi}}\mathrm{erf}(\frac{1}{s_{\theta}})e^{(-1/s_{\theta}^2)}+\frac{s_{\theta}}{2}\sqrt{\frac{2}{\pi}}\mathrm{erf}(\frac{\sqrt{2}}{s_{\theta}}).
\end{equation}

\bibliographystyle{jfm}
\bibliography{jfm}

\end{document}